\documentclass[11pt]{article}

\usepackage[top=50mm, bottom=50mm, left=50mm, right=50mm]{geometry}

\usepackage{amsmath}
\usepackage{amssymb}
\usepackage{amsfonts}
\usepackage{mathtools}
\usepackage{mathrsfs}
\usepackage[scr = dutchcal]{mathalpha}
\usepackage{graphicx}
\usepackage{graphics}
\usepackage{float}
\usepackage{color}
\usepackage{fullpage}
\usepackage[normalem]{ulem} 
\usepackage{makeidx}
\usepackage{xspace}
\usepackage{authblk}
\usepackage{datetime}
\usepackage{microtype}
\usepackage{xpatch}
\usepackage{xstring}
\usepackage{fancyhdr}
\usepackage{mathrsfs}
\usepackage{esint}
\usepackage{upgreek}
\usepackage{siunitx}
\usepackage{chemformula}
\usepackage{tabularx, multirow}
\usepackage{threeparttable}
\usepackage[]{booktabs} 
\usepackage{float}
\usepackage{graphicx,xcolor}
\usepackage[colorlinks=false]{hyperref}
\usepackage[noabbrev]{cleveref}
\usepackage[font = footnotesize, labelfont=bf]{caption}

\fancypagestyle{fpg}{%
	\lfoot{\footnotesize J. Phys. D. \textbf{2021}}%
	\cfoot{\normalsize\thepage}
	\rfoot{\footnotesize \href{}{}}%
}

\makeindex

\begin{document}

\title{\Large\textbf{The~Effects of Process Parameters on Melt-pool Oscillatory Behaviour in Gas~Tungsten~Arc~Welding}}
 
	\author[1,$\dagger$]{Amin Ebrahimi}
	\author[2]{Chris R. Kleijn}
	\author{Marcel J.M. Hermans}
	\author{Ian M. Richardson}
	\affil[1]{\small\textit{Department of Materials Science and Engineering, Faculty of Mechanical, Maritime and Materials~Engineering, Delft University of Technology, Mekelweg 2, 2628~CD~Delft, The~Netherlands}}
	\affil[2]{\small\textit{Department of Chemical Engineering, Faculty of Applied Sciences, Delft University of Technology, van~der~Maasweg~9, 2629~HZ~Delft, The~Netherlands}}
	\affil[$\dagger$]{\textit{Corresponding author, Email: A.Ebrahimi@tudelft.nl}}
	
\date{}
\maketitle
\thispagestyle{fpg}

\begin{abstract}
		Internal flow behaviour and melt-pool surface oscillations during arc welding are complex and not yet fully understood. In~the~present work, high-fidelity numerical simulations are employed to describe the~effects of welding position, sulphur concentration ($60$--\SI{300}{ppm}) and travel speed ($1.25$--$\SI{5}{\milli\meter\per\second}$) on molten metal flow dynamics in fully-penetrated melt-pools. A~wavelet transform is implemented to obtain time-resolved frequency spectra of the~oscillation signals, which overcomes the~shortcomings of the~Fourier~transform in rendering time resolution of the~frequency spectra. Comparing the~results of the~present numerical calculations with available analytical and experimental datasets, the~robustness of the~proposed approach in predicting melt-pool oscillations is demonstrated. The~results reveal that changes in the~surface morphology of the~pool resulting from a~change in welding position alter the~spatial distribution of arc forces and power-density applied to the~molten material, and in turn affect flow patterns in the~pool. Under similar welding conditions, changing the~sulphur concentration affects the~Marangoni flow pattern, and increasing the~travel speed decreases the~size of the~pool and increases the~offset between top and bottom melt-pool surfaces, affecting the~flow structures (vortex formation) on the~surface. Variations in the~internal flow pattern affect the~evolution of melt-pool shape and its surface oscillations.
\end{abstract}

\bigskip
\newpage

\section{Introduction}
\label{sec:intro}

Fusion joining of metallic materials is an essential requirement in many industries. The~integrity of products depends critically on the~joining technique employed and the~quality of the~joints produced~\mbox{\cite{Aucott_2018,DebRoy_1995}}, which in turn is influenced by the~dynamic stability of the~melt-pool~\cite{DebRoy_2018}. A~better understanding of the~complex transport phenomena inside melt-pools offers considerable opportunities for improved monitoring and control of joining processes. To date, control and optimisation of welding processes relies largely on trial-and-error experiments that often pose challenges due to the~non-linearity of melt-pool responses to changes in operating conditions, material properties and process parameters~\cite{Mills_1990,Juang_2002,Cunningham_2019}. Moreover, welding process development requires tolerance to parameter variations, within which the~resultant weld integrity must be fit for the~intended purpose, irrespective of the~particular parameter combinations within the~defined procedural range.

The~present work focuses on positional gas tungsten arc (GTA) welding, which involves several significant operating parameters, the~number increasing when complex time-dependent phenomena are considered~\cite{Lancaster_1986}. The~simulation-based approach utilised in the~present work offers the~potential to reduce procedure development costs and will enhance our understanding of melt-pool behaviour during positional GTA welding. 

The~majority of published studies on GTA melt-pool oscillatory behaviour are experimentally based, consider only the~flat (1G or PA) welding orientation (\textit{i.e.} position C1 shown in \cref{fig:schematic}) and focus on processing the~signals received from the~melt pool to sense penetration~\cite{Li_2017_2}. Experimental techniques employed are often based on laser vision~\cite{Shi_2015}, arc voltage~\cite{Xiao_1990} or arc-light intensity~\cite{Yoo_1993} measurements. A~critical limitation is related to the~inadequate signal-to-noise ratio for low-amplitude surface oscillations~\cite{Tam_1989}, which makes the~application of a~triggering action essential~\cite{Xiao_1992_thesis}. Moreover, these techniques ignore convection in the~melt pool, which is difficult to measure due to opacity, the~fast dynamic response of the~molten metal flow and high temperatures~\cite{Aucott_2018}. In addition to the~experimental measurements, analytical models have been developed to predict dominant oscillation frequencies. These models are based on similarities between oscillations of the~melt-pool surface and the~vibrations of a~thin stretched membrane~\cite{Xiao_1990,Xiao_1993,Andersen_1997,Yoo_1993,Maruo_1993}. Unfortunately, the~absolute accuracy of these analytical models is critically dependent on the~melt-pool shape, temperature-dependent material properties and processing conditions~\cite{Tam_1989}, which in turn are affected by unsteady transport phenomena in the~pool~\cite{Wu_2020}; factors that are not known \textit{a priori}. Moreover, changes in oscillation mode and amplitude are not predictable using these models. Conversely, high-fidelity numerical simulations have demonstrated a~remarkable potential to describe the~complex internal flow behaviour in melt pools and associated surface deformations~\cite{Ebrahimi_2021,Cook_2020}.

Although many numerical models are available (\textit{e.g.}~\cite{Kou_1985,Zacharia_1989,Wu_1994,Wu_2007,Mishra_2008,Traidia_2011,Mougenot_2013,Hao_2020}), numerical studies on melt-pool surface oscillations are scarce and the~melt-pool oscillatory behaviour is not yet fully understood, particularly for positional welding conditions. Previous studies often focused on the~influence of surface deformations on the~melt-pool shape~\cite{Thompson_1989,Tsai_1989,Kim_1992,Zhang_1996,Cao_1998} or the~morphology of the~melt-pool surface, to study ripple formation~\cite{Liu_2015}, welding defects such as undercut~\cite{Meng_2016_2} or humping~\cite{Feng_2020,Du_2019,Pan_2016}. Chen~\textit{et al.}~\cite{Chen_1998} revealed that the~flow patterns in pools with one free surface, representative of partially penetrated welds, differ from those in pools with two free surfaces, representative of fully penetrated welds. In their model, they neglected solidification and melting, assumed that the~flow inside the~pool is axisymmetric and that the~surface tension of the~molten material is a~linear function of temperature. Using a~similar model, including electromagnetic forces and solid-liquid phase transformations, \hbox{Ko~\textit{et al.}~\cite{Ko_2000,Ko_2001}} showed that pool oscillations during stationary GTA welding depend on the~direction of the~Marangoni flow. For many real-world welding applications, non-pure materials are involved for which the~surface tension changes non-linearly with temperature~\cite{Sahoo_1988}. This coupled with movements of the~melt-pool boundaries due to solid-liquid phase transformations form complex unsteady flow patterns that are inherently three-dimensional~\cite{Joshi_1997,Zhao_2009,Kidess_2016}, affecting the~melt-pool surface oscillations. 

In our previous study on stationary GTA welding~\cite{Ebrahimi_2021}, we demonstrated that changes in welding current and material properties can alter the~time-frequency response of melt-pool oscillations under both partial and full penetration conditions. The~welding position can also influence the~shape of the~deformations of the melt-pool surface and its shape during GTA welding~\cite{Kang_2003,Nguyen_2017}. Moreover, morphology of the~melt-pool surface can affect the~spatial and temporal distribution of arc-pressure and power-density and thus the~melt-pool dynamic behaviour~\cite{Ko_2000,Ebrahimi_2021}. Further investigations are essential to broaden our understanding of complex internal flow behaviour in melt pools and melt-pool surface oscillations in positional GTA welding. In the~present study, the~results of numerical simulations employed to reveal complex unsteady transport phenomena in the~melt-pool and associated surface oscillations during positional GTA welding are reported. The~present study focuses particularly on fully-penetrated pools, where the~melt-pool oscillations are critical for process stability; however, the~present model is equally applicable to partial penetration conditions. The~coupling between melt-pool surface deformations, arc force and power-density distributions, which represent physical realism and can affect the~predicted thermal and flow fields, are taken into account. The~continuous wavelet transform is applied to the~time-resolved displacement signals acquired from the~simulations to enhance our understanding of the~evolution of surface oscillations and its correlation with process parameters and material properties. A~novel insight into the~evolution of melt-pool surface oscillations and the~complex flow inside molten metal melt pools is provided, which offers a~computational approach to fusion welding process development and optimisation.

\section{Problem description}
\label{sec:problem_des}

\begin{figure}[H] 
	\centering
	\includegraphics[width=1.0\linewidth]{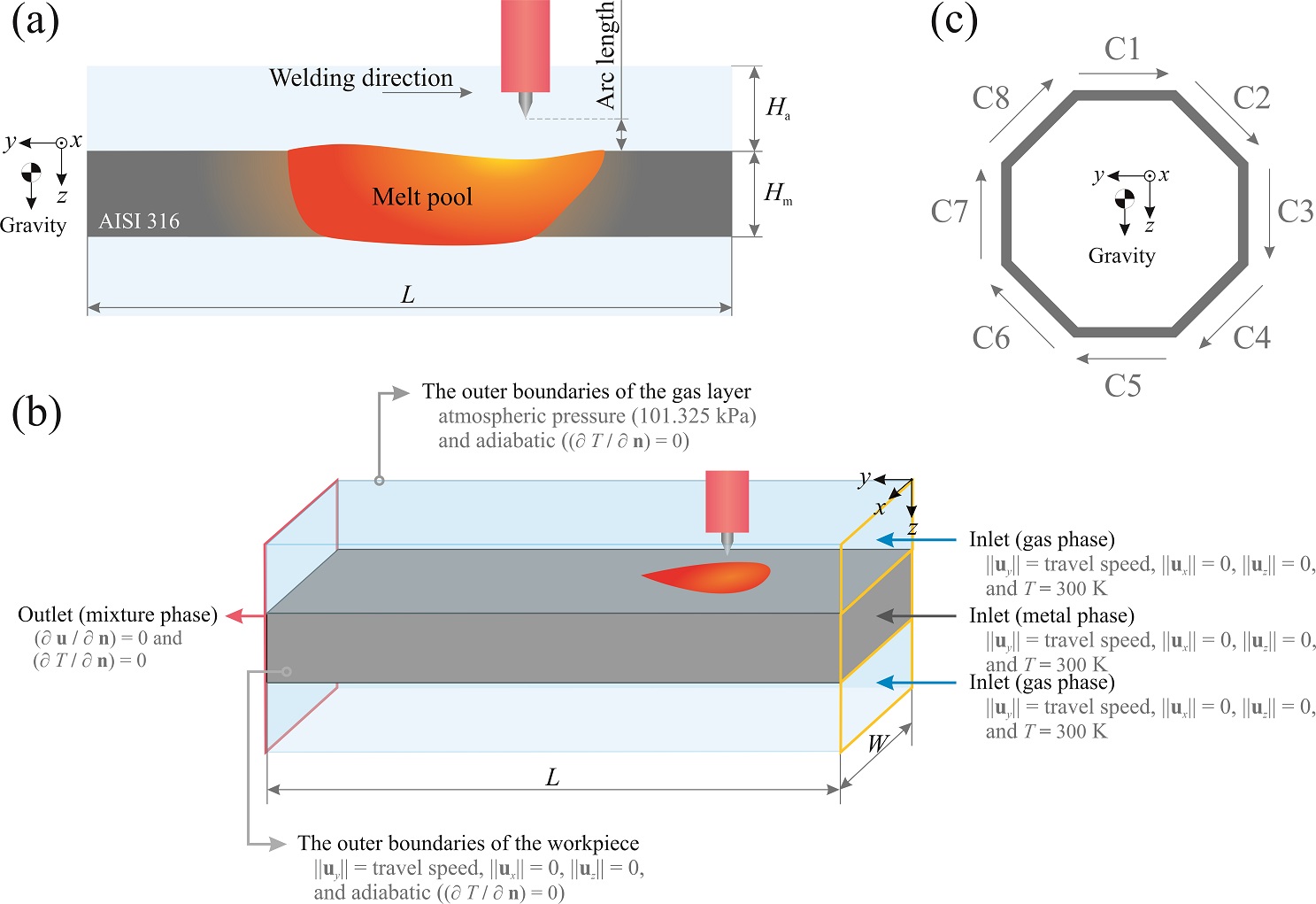}
	\caption{Schematic of moving gas tungsten arc welding (GTAW). (a)~a~cross section of the~plate showing a~fully-penetrated melt pool, (b)~three-dimensional view of the~problem under consideration and (c)~different welding positions studied in the~present work.}
	\label{fig:schematic}
\end{figure}

Molten metal flow behaviour and associated surface oscillations during positional gas tungsten arc welding of a~stainless steel (AISI~316) plate are studied numerically. As shown schematically in \cref{fig:schematic}, the~plate has a~thickness of $H_\mathrm{m} = \SI{2}{\milli\meter}$ and is heated locally by an electric arc plasma to create a~melt pool in the~plate that is initially at $T_\mathrm{0} = \SI{300}{\kelvin}$. A~perpendicular torch is adopted for all positions in line with automated pipeline welding. The~current is set to $\SI{85}{\ampere}$ and the~initial arc length (electrode tip to workpiece distance) before igniting the~arc is $\SI{2.5}{\milli\meter}$. Obviously, the~melt-pool surface deforms during the~process, leading to changes in the~length, voltage and power of the~arc as well as power-density distribution and the~magnitude and distribution of forces induced by the~arc plasma~\cite{Tsai_1985,Lin_1986,Ebrahimi_2021}. In~the~present model, the~melt-pool is decoupled from the~arc plasma to reduce the~computation time and complexity of simulations. Accordingly, the~related source terms for momentum and thermal energy are adjusted dynamically in the~present model to take these changes into account, as explained \cref{sec:methods}. Argon gas is employed to shield the~melt-pool. The~temperature-dependent thermophysical properties of AISI~316 and argon are presented in \cref{tab:material_properties}. The~temperature-dependent surface tension of the~molten material is modelled using an~empirical correlation introduced in \cref{eq:surface_tension}~\mbox{Sahoo~\textit{et al.}~\cite{Sahoo_1988}}, which accounts for the~influence of surface-active elements (\textit{i.e.} sulphur). 

A moving reference frame is employed in the~present numerical simulations to simulate unsteady convection in the~melt pool. Hence, instead of moving the~heat source, the~material enters the~computational domain, translates at a~fixed speed (\textit{i.e.} the~welding travel speed) opposite to the~welding direction shown in \cref{fig:schematic}, and leaves the~computational domain. Applying the~moving reference frame technique facilitates a~decrease in the~size of the~computational domain and thus the~runtime. The~computational domain is designed in the form of a~rectangular cuboid encompassing the~workpiece and two layers of gas below and above the~sample to monitor the~oscillations of the~melt-pool surfaces. The~width and the~length of the~computational domain is $W = \SI{40}{\milli\meter}$ and $L = \SI{70}{\milli\meter}$ respectively, which is substantially larger than the~dimensions of the~melt-pool. The~boundary conditions applied to the computational domain in the~simulations are shown in \cref{fig:schematic}(b). The~outer boundaries of the~workpiece are adiabatic and are treated as no-slip moving walls. The~gas layers have a~thickness of $H_\mathrm{a} = \SI{2}{\milli\meter}$ and a~fixed atmospheric pressure ($p = p_\mathrm{atm} = \SI{101.325}{\kilo\pascal}$) is applied to their outer boundaries. The~electrode axis is located in the~middle of the~plate (\textit{i.e.}~$x = W/2$) and $\SI{15}{\milli\meter}$ away from the~leading-edge of the~plate (\textit{i.e.} $y = \SI{15}{\milli\meter}$). Eight different welding positions (workpiece orientations with respect to gravity) are studied, as shown in \cref{fig:schematic}(c).

\bgroup
\def\arraystretch{1.15}	
\begin{table}[H] 
	\centering
	\caption{Thermophysical properties of AISI~316 and argon used in the~computational model. $T$ is in Kelvin.}
	\begin{tabular}{llll}
		\toprule
		Property                              & Stainless steel (AISI~316)~\cite{Mills_2002}                         & Gas (argon)  & Unit                                \\ \midrule
		Density $\rho$                        & \SI{7100}                                        & \SI{1.623}   & \si{\kilogram\per\meter\cubed}      \\
		Specific heat capacity $c_\mathrm{p}$ & $\SI{430.18} + \SI{0.1792}\cdot T$ (solid phase) & \SI{520.64}  & \si{\joule\per\kilogram\per\kelvin} \\
		& \SI{830} (liquid phase)                          &              &                                     \\
		Thermal conductivity $k$              & $\SI{11.791} + \SI{0.0131}\cdot T$ (solid phase) & \SI{520.64}  & \si{\watt\per\meter\per\kelvin}     \\
		& $\SI{6.49} + \SI{0.0129}\cdot T$ (liquid phase)  &              &                                     \\ 
		Viscosity $\mu$                       & \SI{6.42e-3}                                     & \SI{2.1e-05} & \si{\kilogram\per\meter\per\second} \\
		Thermal expansion coefficient $\beta$ & \SI{2e-06}                                       & --           & \si{\per\kelvin}                    \\
		Latent heat of fusion $\mathcal{L}$   & \SI{2.7e5}                                       & --           & \si{\joule\per\kilogram}            \\
		Liquidus temperature $T_\mathrm{l}$   & \SI{1723}                                        & --           & \si{\kelvin}                        \\
		Solidus temperature $T_\mathrm{s}$    & \SI{1658}                                        & --           & \si{\kelvin}                        \\ \bottomrule
	\end{tabular} 
	\label{tab:material_properties}
\end{table}
\egroup

\section{Methods} 
\label{sec:methods}

\subsection{Model formulation}
\label{sec:math_model}	

Our previous multiphase model for stationary GTA welding \cite{Ebrahimi_2021} is extended here to predict the~complex three-dimensional molten metal flow and melt-pool surface oscillations during positional moving GTA welding. In this model, the~molten metal and argon were assumed to be incompressible and were treated as Newtonian fluids. Accordingly, the~unsteady governing equations were cast in conservative form as follow:

\begin{equation}
	\nabla\cdot\mathbf{u} = 0,
	\label{eq:mass}
\end{equation}

\begin{equation}
	\rho \frac{D \mathbf{u}}{D t} = \mu \nabla^2 \mathbf{u} -\nabla p + \mathbf{F}_\mathrm{d} + \mathbf{F}_\mathrm{s} + \mathbf{F}_\mathrm{b},
	\label{eq:momentum}
\end{equation}

\begin{equation}
	\rho \frac{D h}{D t} = \frac{k}{c_\mathrm{p}} \nabla^2 h - \rho \frac{D \left(\psi \mathcal{L}\right)}{D t} + S_\mathrm{q} + S_\mathrm{l},
	\label{eq:energy}
\end{equation}	

\noindent
where, $\rho$ is the~density, $\mathbf{u}$ the~relative velocity vector, $t$ the~time, $\mu$ the~dynamic viscosity, $p$ the~pressure, $h$ the~sensible heat, $k$ the~thermal conductivity, $c_\mathrm{p}$ the~specific heat capacity at constant pressure, and $\left(\psi \mathcal{L}\right)$ the~latent~heat. The~total enthalpy of the~material $\mathscr{H}$ is the~sum of the~latent heat $\left(\psi \mathcal{L}\right)$ and the~sensible heat $h$ and is defined as follows~\cite{Voller_1991}:

\begin{equation}
	\mathscr{H} = \left(h_\mathrm{r} + \int_{T_\mathrm{r}}^{T} c_\mathrm{p} \mathop{}\!\mathrm{d}T\right) + \psi \mathcal{L},
	\label{eq:enthalpy}
\end{equation}

\noindent
where, $T$ is the~temperature, $\psi$ the~local liquid volume-fraction, and $\mathcal{L}$ the~latent heat of fusion. The~subscript `r' indicates the~reference condition. Assuming the~liquid volume-fraction $\psi$ to be a~linear function of temperature \cite{Voller_1991}, its value can be calculated as follows:

\begin{equation}
	\psi = \frac{T - T_\mathrm{s}}{T_\mathrm{l} - T_\mathrm{s}}; \quad T_\mathrm{s} \le T \le T_\mathrm{l},
	\label{eq:liquid_fraction}
\end{equation}

\noindent
where, $T_\mathrm{l}$ and $T_\mathrm{s}$ are the~liquidus and solidus temperatures, respectively.

To capture the~position of the~gas-metal interface, the~volume-of-fluid (VOF) method \cite{Hirt_1981} was employed, where the~scalar function $\phi$ indicates the~local volume-fraction of a~phase in a given computational cell. The~value of $\phi$ varies from 0 in the~gas phase to 1 in the~metal phase, and cells with $0 < \phi < 1$ represent the~gas-metal interface. The~linear advection equation describes the~advection of the~scalar function $\phi$ as follows: 

\begin{equation}
	\frac{D \phi}{D t} + \phi \nabla\cdot\mathbf{u} = 0.
	\label{eq:vof}
\end{equation}

\noindent
Accordingly, the~effective thermophysical properties of the~material in each computational cell were determined as follows:

\begin{equation}
	\xi = \phi \, \xi_\mathrm{m} + \left(1-\phi\right) \xi_\mathrm{g},
	\label{eq:mixture_model}
\end{equation}

\noindent
where, $\xi$ corresponds to thermal conductivity $k$, specific heat capacity $c_\mathrm{p}$, viscosity $\mu$ and density $\rho$, and subscripts `g' and `m' indicate gas or metal respectively.

Solid-liquid phase transformation occurs in the~temperature range between $T_\mathrm{s}$ and $T_\mathrm{l}$ in the~so-called `mushy zone'. To model the~damping of liquid velocities in the~mushy zone, and suppression of liquid velocities in solid regions, the~sink term $\mathbf{F}_\mathrm{d}$ based on the~enthalpy-porosity technique \cite{Voller_1987}, was incorporated into the~momentum equation and is defined as

\begin{equation}
	\mathbf{F}_\mathrm{d} = C\ \frac{(1 - \psi)^2}{\psi^3 + \epsilon} \ \mathbf{u},
	\label{eq:sink_term}
\end{equation}

\noindent
where, $C$ is the~mushy-zone constant and its value was chosen to equal $10^7\,\SI{}{\kilogram\per\square\meter\per\square\second}$, in accordance with \cite{Ebrahimi_2019} to avoid numerical artefacts associated with inappropriate assignment of this parameter, and $\epsilon$ is a~constant equal to $10^{-3}$ employed to avoid division by zero. 

To apply forces on the~gas-metal interface, the~continuum surface force (CSF) model~\cite{Brackbill_1992} was employed. In~the~CSF model, surface forces are considered as volumetric forces acting on the~material contained in grid cells in the~interface region. The~source term $\mathbf{F}_\mathrm{s}$ was added to \cref{eq:momentum} as follows:

\begin{equation}
	\mathbf{F}_\mathrm{s} = \mathbf{f}_\mathrm{s} \lVert \nabla\phi \rVert \frac{2\rho}{\rho_\mathrm{m} + \rho_\mathrm{g}},
	\label{eq:CSF_model}
\end{equation}

\noindent
where, the~surface force applied to a unit area $\mathbf{f}_\mathrm{s}$ comprises arc plasma, surface tension and Marangoni forces and was defined as follows:

\begin{equation}
	\mathbf{f}_\mathrm{s} = \mathbf{f}_\mathrm{a} + \gamma \kappa \hat{\mathbf{n}} + \frac{\mathrm{d} \gamma}{\mathrm{d} T} \left[\nabla T - \hat{\mathbf{n}}\left(\hat{\mathbf{n}} \cdot \nabla T\right)\right],
	\label{eq:surface_force}
\end{equation}

\noindent
where, $\mathbf{f}_\mathrm{a}$ is arc plasma force, $\gamma$ the~surface tension, $\hat{\mathbf{n}}$ the~surface unit normal vector $\left(\hat{\mathbf{n}} = \nabla\phi / \lVert \nabla\phi \rVert \right)$ and $\kappa$ the~surface curvature $\left(\kappa = \nabla\cdot\hat{\mathbf{n}}\right)$.

The~arc plasma force $\mathbf{f}_\mathrm{a}$ defined in \cref{eq:surface_force} comprises arc plasma shear stress $\mathbf{f}_\tau$ and arc pressure $\mathbf{f}_\mathrm{p}$,

\begin{equation}
	\mathbf{f}_\mathrm{s} = \mathbf{f}_\tau + \mathbf{f}_\mathrm{p}.
	\label{eq:arc_force}
\end{equation}

\noindent
The~arc plasma shear stress $\mathbf{f}_\tau$, which applies tangent to the~surface, was defined as follows~\cite{Bai_2018}:

\begin{equation}
	\mathbf{f}_\tau = \left[\tau_\mathrm{max} \: g_\tau\left(\mathscr{R}, \sigma_\tau \right)\right] \hat{\mathbf{t}},
	\label{eq:arc_shear_force}
\end{equation}

\noindent
where, the~maximum arc shear stress $\tau_\mathrm{max}$ \cite{Osti_1996,Lee_1995}, the~arc shear stress distribution function $g_\tau$ \cite{Unnikrishnakurup_2017} and the~surface unit tangent vector $\hat{\mathbf{t}}$ \cite{Bai_2018} were defined as follows:

\begin{equation}
	\tau_\mathrm{max} = \SI{7e-2}{} I^{1.5} \exp\left(\frac{\SI{-2.5e4}{} \bar{\ell}}{I^{0.985}} \right),
	\label{eq:max_shear_force}
\end{equation}

\begin{equation}
	g_\tau\left(\mathscr{R}, \sigma_\tau \right) = \sqrt{\frac{\mathscr{R}}{{\sigma_\tau}}} \exp\left(\frac{-\mathscr{R}^2}{{\sigma_\tau}^2}\right) ,
	\label{eq:dist_shear_force}
\end{equation}

\begin{equation}
	\hat{\mathbf{t}} = \frac{\mathbf{r} - \hat{\mathbf{n}} \left(\hat{\mathbf{n}} \cdot \mathbf{r}\right)}{\lVert \mathbf{r} - \hat{\mathbf{n}} \left(\hat{\mathbf{n}} \cdot \mathbf{r}\right) \rVert}.
	\label{eq:tangent_vec}
\end{equation}

\noindent
Here, $I$ is the~welding current in Amperes, $\bar{\ell}$ is the~mean arc length in meters, $\mathscr{R}$~the~radius in $x$-$y$ plane~(\textit{i.e.}~$\mathscr{R} = \sqrt{x^2 + y^2}$) in meters, and $\mathbf{r}$ the~position vector in the~$x$-$y$ plane in meters. The~distribution parameter $\sigma_\tau$ (in~meters) is assumed to be a~function of the~mean arc length $\bar{\ell}$ and current $I$ and was approximated on the basis of the~data reported by Lee and Na~\cite{Osti_1996}:

\begin{equation}
	\sigma_\tau = \SI{1.387e-3}{} + I^{-0.595} \bar{\ell}^{0.733}.
	\label{eq:dist_shear_stress}
\end{equation}

\noindent
The~arc pressure $\mathbf{f}_\mathrm{p}$ was determined as follows \cite{Lin_1986}:

\begin{equation}
	\mathbf{f}_\mathrm{p} = \mathscr{F}_\mathrm{p} \left[\frac{\mu_0 I}{4 \pi} \frac{I}{2 \pi {\sigma_\mathrm{p}}^2} \exp\left(\frac{-\mathscr{R}^2}{2 {\sigma_\mathrm{p}}^2}\right) \right] \hat{\mathbf{n}},
	\label{eq:arc_pressure}
\end{equation}

\noindent
where, $I$ is the~current in Amperes, and $\mu_0$ is the~vacuum permeability equal to $4\pi\cdot10^{-7}\,\SI{}{\henry\per\meter}$. The~distribution parameter $\sigma_\mathrm{p}$ (in metres) was determined using the~experimental data reported by Tsai and Eagar~\cite{Tsai_1985} for an~argon arc with an~electrode tip angle of $75^\circ$ as follows:

\begin{equation}
	\sigma_\mathrm{p} = \SI{7.03e-2}{} \ell^{0.823} + \SI{2.04e-4}{} I^{0.376},
	\label{eq:dist_pressure}
\end{equation}

\noindent
where, $\ell$ is the~local arc length in meters, and $I$ the~current in Amperes. Hence, spatial and temporal variations of the~arc pressure distribution resulted from changes in morphology of the~melt-pool surface were taken into account. Changes in surface morphology can cause the~total arc force applied to the melt-pool surface~($\iiint \limits_{\forall} \lVert \mathbf{f}_\mathrm{p} \rVert \mathop{}\!\mathrm{d}V$) to differ from the~expected arc force ($\mu_0 I^2 / 4 \pi$) due to changes in $\lVert \nabla\phi \rVert$~\cite{Ebrahimi_2021,Meng_2016}. This numerical artefact was negated by incorporating $\mathscr{F}_\mathrm{p}$, which is defined as follows:

\begin{equation}
	\mathscr{F}_\mathrm{p} = \alpha \, \frac{\mu_0 I^2}{4 \pi} \frac{1}{\iiint \limits_{\forall} \lVert \mathbf{f}_\mathrm{p} \rVert \mathop{}\!\mathrm{d}V}.
	\label{eq:adj_pressure}
\end{equation}

\noindent
The~dimensionless factor $\alpha$ was employed, as suggested by Liu~\textit{et al.}~\cite{Liu_2015} and Lin and Eagar~\cite{Lin_1986}, to match the~theoretically determined arc pressure with experimentally measured values , and was calculated as follows:

\begin{equation}
	\alpha = 3 + \SI{8e-3}{} I,
	\label{eq:adj_pressure_coeff}
\end{equation}

\noindent
with $I$ the~welding current in Amperes.

\begin{figure}[H] 
	\centering
	\includegraphics[width=0.50\linewidth]{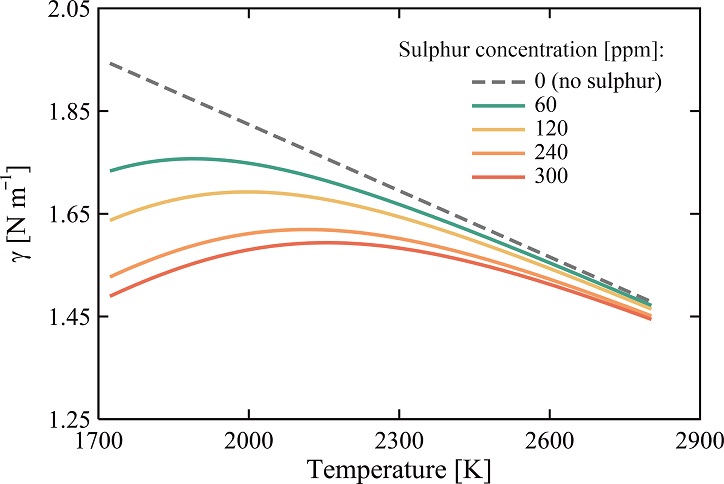}
	\caption{Surface tension of a molten Fe-S alloy as a function of temperature approximated using \cref{eq:surface_tension} for different sulphur concentrations.}
	\label{fig:surface_tension}
\end{figure}

The~temperature-dependent surface tension of the~molten material was modelled using an~empirical correlation~\cite{Sahoo_1988} that accounts for the~influence of sulphur as a surface-active element, and is defined as follows:

\begin{equation}
	\gamma = \gamma_\mathrm{m}^\circ + \left(\frac{\partial\gamma}{\partial T}\right)^\circ \left(T - T_\mathrm{m}\right) - \mathrm{R}\, T\, \Gamma_\mathrm{s}\, \ln\!\left[1 + \psi\, a_\mathrm{s} \exp\!\left(\frac{-\Delta H^\circ}{\mathrm{R} T}\right)\right],
	\label{eq:surface_tension}
\end{equation}

\noindent
where, $\gamma_\mathrm{m}^\circ$ is the~surface tension of the~pure molten-material at the~melting temperature $T_\mathrm{m}$, $\left(\partial \gamma / \partial T\right)^\circ$ the~temperature gradient of the~surface tension of the~pure molten-material, $T$ the~temperature in Kelvin, $\Gamma_\mathrm{s}$ the~adsorption at saturation, $\mathrm{R}$ the~universal gas constant, $\psi$ an entropy factor, $a_\mathrm{s}$ the~activity of the~solute, and $\Delta H^\circ$ the~standard heat of adsorption. The~values reported by Sahoo~\textit{et al.}~\cite{Sahoo_1988} were used for the~properties in \cref{eq:surface_tension}. Variations of the~surface tension of a molten Fe-S alloy with temperature are shown in \cref{fig:surface_tension} for different sulphur concentrations studied in the~present work. It should be noted that in this model it is assumed that sulphur is distributed uniformly over the~melt-pool surface. Since temperatures in the~melt-pool are below the boiling temperature of stainless steel ($\mathscr{O}(\SI{3000}{\kelvin})$) and there is no material addition with a~different sulphur concentration, the~effects of changes in the~sulphur concentration is virtually negligible for the~cases studied in the~present work. Therefore, the~effects of adsorption, desorption and redistribution of species along the~melt-pool surface are neglected in the~present numerical simulations.

$\mathbf{F}_\mathrm{b}$ in \cref{eq:momentum} is the~body force, which comprises electromagnetic, gravity and thermal buoyancy forces. The~electromagnetic force was computed using the~model proposed by~Tsao and Wu~\cite{Tsao_1988} transformed into a~body-fitted coordinate system, and the~thermal buoyancy force was modelled using the~Boussinesq approximation \cite{Tritton_1977}. Hence, the~body forces are defined as follows:

\begin{equation}
	{\mathbf{f}_\mathrm{b}}_x = \frac{- \mu_0 I^2}{4\pi^2 {\sigma_\mathrm{e}}^2 \mathscr{R}} \exp\left(\frac{- \mathscr{R}^2}{2{\sigma_\mathrm{e}}^2}\right) \left[1 - \exp\left(\frac{- \mathscr{R}^2}{2{\sigma_\mathrm{e}}^2}\right)\right] \left(1 - \frac{z - z^\prime}{H_\mathrm{m} - z^\prime}\right)^2 \left(\frac{x}{\mathscr{R}}\right),
	\label{eq:body_force_x}
\end{equation}

\begin{equation}
	{\mathbf{f}_\mathrm{b}}_y = \frac{- \mu_0 I^2}{4\pi^2 {\sigma_\mathrm{e}}^2 \mathscr{R}} \exp\left(\frac{- \mathscr{R}^2}{2{\sigma_\mathrm{e}}^2}\right) \left[1 - \exp\left(\frac{- \mathscr{R}^2}{2{\sigma_\mathrm{e}}^2}\right)\right] \left(1 - \frac{z - z^\prime}{H_\mathrm{m} - z^\prime}\right)^2 \left(\frac{y}{\mathscr{R}}\right),
	\label{eq:body_force_y}
\end{equation}

\begin{equation}
	{\mathbf{f}_\mathrm{b}}_z = \frac{- \mu_0 I^2}{4\pi^2 {\sigma_\mathrm{e}}^2 \mathscr{R}} \exp\left(\frac{- \mathscr{R}^2}{2{\sigma_\mathrm{e}}^2}\right) \left(1 - \frac{z - z^\prime}{H_\mathrm{m} - z^\prime}\right) + \rho \mathbf{g} - \rho \beta \left(T - T_\mathrm{l}\right) \mathbf{g}.
	\label{eq:body_force_z}
\end{equation}

\noindent
Here, the~distribution parameter for the~electromagnetic force $\sigma_\mathrm{e}$ is the~same as $\sigma_\mathrm{p}$, according to~\mbox{Tsai and Eagar~\cite{Tsai_1985}}, $z^\prime$~is the~position of the~melt-pool surface in $x$-$y$ plane, $\beta$ the~thermal expansion coefficient, and $\mathbf{g}$ the~gravitational acceleration vector.

The~thermal energy provided by the~arc was modelled by adding the~source term $S_\mathrm{q}$ to the~energy equation (\cref{eq:energy}) and was defined as

\begin{equation}
	S_\mathrm{q} = \mathscr{F}_\mathrm{q} \left[\frac{\eta I U}{2 \pi {\sigma_\mathrm{q}}^2} \exp\left(\frac{-\mathscr{R}^2}{2 {\sigma_\mathrm{q}}^2}\right) \lVert \nabla\phi \rVert \frac{2\, \rho \, c_\mathrm{p}}{(\rho \, c_\mathrm{p})_\mathrm{m} + (\rho \, c_\mathrm{p})_\mathrm{g}} \right],
	\label{eq:arc_heat}
\end{equation}

\noindent
where, the~process efficiency $\eta$ considered to be a~linear function of welding current, varying from $80\%$ at $\SI{50}{\ampere}$ to $70\%$ at $\SI{300}{\ampere}$~\cite{Murphy_2018}. It should be noted that the~source term $S_\mathrm{q}$ is only applied to the~top surface of the~workpiece. The~arc voltage $U$ depends on welding current and arc length, and was determined as follows:

\begin{equation}
	U = U_\mathrm{o} + U_\mathrm{I} I + U_\mathrm{e} \bar{\ell},
	\label{eq:voltage_variations}
\end{equation}

\noindent
where, $U_\mathrm{o}$ is the~electrode fall voltage equal to $\SI{8}{\volt}$ \cite{Richardson_1991}, $U_\mathrm{I}$ the~coefficient of variation of arc voltage with current equal to $\SI{1.3e-2}{\volt\per\ampere}$ \cite{Goodarzi_1997}, and $U_\mathrm{e}$ the~electric field strength equal to $\SI{7.5}{\volt\per\centi\meter}$~\cite{Richardson_1991}. Using the~data reported by~Tsai and Eagar~\cite{Tsai_1985}, the~distribution parameter $\sigma_\mathrm{q}$ (in meters) was determined as follows:

\begin{equation}
	\sigma_\mathrm{q} = \SI{1.61e-1}{} \ell^{0.976} + \SI{2.23e-4}{} I^{0.395},
	\label{eq:dist_heat}
\end{equation}

\noindent
with $\ell$ in meters and $I$ in Amperes. The~adjustment factor $\mathscr{F}_\mathrm{q}$ was used to negate changes in the~total heat input due to surface deformations, which is defined as follows:

\begin{equation}
	\mathscr{F}_\mathrm{q} = \frac{\eta I U}{\iiint \limits_{\forall} S_\mathrm{q} \mathop{}\!\mathrm{d}V}.
	\label{eq:adj_heat}
\end{equation}

The~sink term $S_\mathrm{l}$ was added to the~energy equation to account for heat losses due to convection and radiation, and was determined as follows:

\begin{equation}
	S_\mathrm{l} = - \left[\mathscr{h}_\mathrm{c} \left(T - {T_\mathrm{0}}\right) + \mathcal{K}_\mathrm{b} \mathscr{E} \left(T^4 - {T_\mathrm{0}}^4\right)\right] \lVert \nabla\phi \rVert \frac{2\, \rho \, c_\mathrm{p}}{(\rho \, c_\mathrm{p})_\mathrm{m} + (\rho \, c_\mathrm{p})_\mathrm{g}},
	\label{eq:heat_loss}
\end{equation}

\noindent
where, $\mathscr{h}_\mathrm{c}$ is the~heat transfer coefficient equal to $\SI{25}{\watt\per\square\meter\per\kelvin}$ \cite{Johnson_2017}, $\mathcal{K}_\mathrm{b}$ the~Stefan--Boltzmann constant and $\mathscr{E}$~the~radiation emissivity equal to $0.45$ \cite{Sridharan_2011}.

\subsection{Numerical implementation}
\label{sec:num_proc}

The~numerical simulations reported in the~present work make use of a~finite-volume solver, ANSYS~Fluent~\cite{Ansys}. The~surface-tension model as well as the~source terms in the~governing equations were implemented in the~solver using user-defined functions programmed in the~C programming language. The~computational grid contains about~$\SI{5.2e6}{}$ hexahedral cells, where cell spacing varies gradually from $\SI{40}{\micro\meter}$ in the~melt-pool region and close to the~gas-metal interface to~$\SI{400}{\micro\meter}$ close to the~boundaries of the~computational domain. Spatial discretisation was implemented by the~central-differencing scheme for momentum advection and diffusive fluxes, and the~time marching was performed employing a~first order implicit scheme. The~time-step size was set to~$10^{-5}\, \SI{}{\second}$, resulting in a~Courant number $(\mathrm{Co} = \lVert \mathbf{u} \rVert \Delta t / \Delta x)$ less than $0.25$. Velocity and pressure fields were coupled using the~PISO~(pressure-implicit with splitting of operators) scheme \cite{Issa_1986} and the~pressure interpolation was performed employing the~PRESTO~(pressure staggering option) scheme \cite{Patankar_1980}. The~advection of the~volume-fraction field was formulated using an~explicit compressive VOF method~\cite{Ubbink_1997}. Convergence for each time-step is achieved when scaled residuals fall below~$10^{-7}$. Each simulation was executed in parallel on $80$ cores~(AMD~EPYC~7452) of a~computing cluster for a~total run-time of about $\SI{800}{\hour}$. To~reduce the~computational costs associated with running the~model, possible developments can focus on performing model order reduction or decreasing the~spatial and temporal resolutions of the~simulations. Reliability, validity and grid independence of the~present numerical model in predicting internal flow behaviour, evolution of the~melt-pool shape and surface oscillations were meticulously verified in our previous works~\cite{Ebrahimi_2019_conf,Ebrahimi_2019,Ebrahimi_2020,Ebrahimi_2021}. Moreover, the~frequencies acquired from the~present three-dimensional numerical simulations deviate less than~10\% from the~experimental data reported by~\mbox{Yudodibroto~\cite{Yudodibroto_2010_thesis}}.

\subsection{Time-Frequency analysis}
\label{sec:time_frequency}	

Because of the~complex unsteady molten metal flow in the~melt pool, the~frequency spectra of signals received from an oscillating melt-pool during GTA welding are often time-variant. The~dynamic features of the~oscillation signals cannot be disclosed employing the~conventional Fourier transform~(FT) analysis~\cite{Rioul_1991}. The~continuous wavelet transform (CWT)~\cite{Mallat_2009} was employed in the~present study to overcome the~shortcomings of the~conventional fast Fourier transform (FFT) analysis in characterising the~non-stationary features of the~signals that may contain abrupt changes in their frequency spectra. Employing the~CWT method, the~time-resolved melt-pool surface oscillation signals obtained from the~numerical simulations can be decomposed into time and frequency spaces simultaneously. The~principle of signal processing based on the~wavelet transform is described in \cite{Mallat_2009} and is not repeated here. The~Python programming language was employed to perform time-frequency analysis using the~oscillation signals acquired at a~sampling frequency of~$10^{5}\, \SI{}{\hertz}$. The~Morlet wavelet function, which is a~Gaussian-windowed complex sinusoid, was used as the~mother wavelet that yields an adequate balance in both time and frequency domains.

\section{Results}
\label{sec:results}

\begin{figure}[H] 
	\centering
	\includegraphics[width=0.9\linewidth]{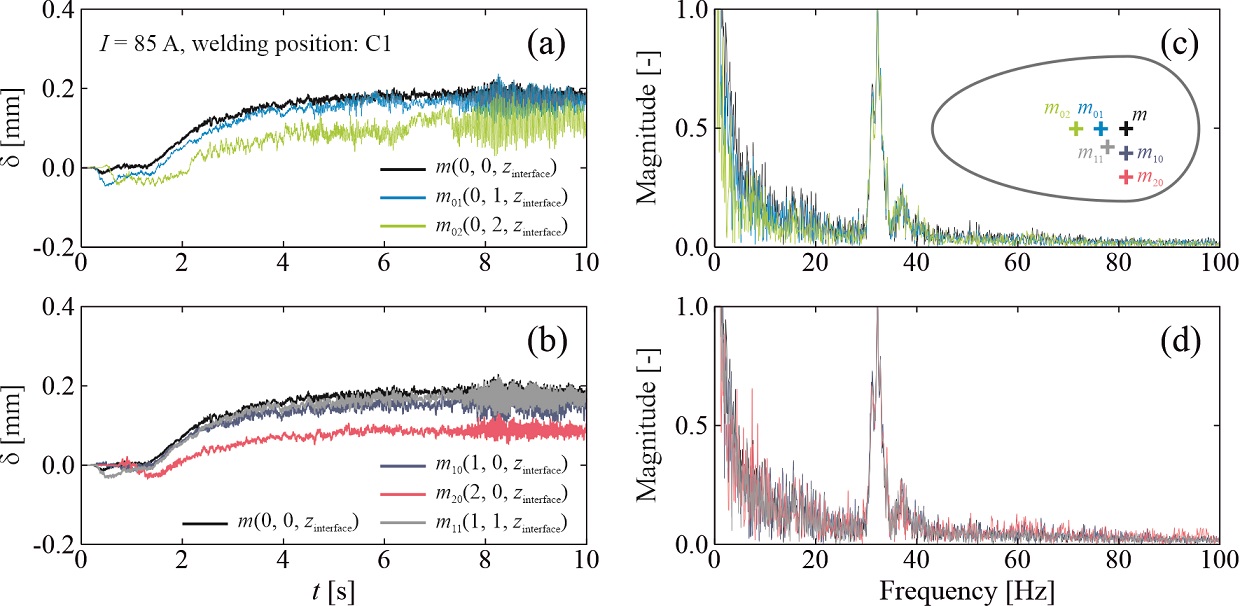}
	\caption{The~displacement signals recorded from the~monitoring points on the~melt-pool top surface and the~corresponding frequency spectra for welding position C1. (a and b)~displacement signals obtained from the~simulations, (c and d)~frequency spectra obtained from FFT analysis. Magnitudes in frequency spectrum are normalised with respect to the~maximum magnitude. Positive values of $\delta$ indicate surface depression and its negative values indicate surface elevation. ($I = \SI{85}{\ampere}$, travel~speed:~$\SI{2.5}{\milli\meter\per\second}$ and sulphur~concentration:~$\SI{240}{ppm}$)}
	\label{fig:surface_deformation_monitor}
\end{figure}

In this section, the~effects of different welding positions (shown in \cref{fig:schematic}(c)) as well as travel speed (ranging between \SI{1.25}{\milli\meter\per\second} and \SI{5}{\milli\meter\per\second}) and sulphur concentration (ranging between \SI{60}{ppm} and \SI{300}{ppm}) on internal flow behaviour, evolution of the~melt-pool shape and surface oscillations during GTA~welding are described. Displacement signals required for characterising the~molten metal oscillations were recorded from several monitoring points distributed over the~melt-pool surface during the~simulations, and are shown in~\cref{fig:surface_deformation_monitor} for welding position C1. Although the~amplitudes of oscillations are different at different locations, the~frequency spectra obtained from FFT analysis look similar. Hence, the~signals recorded from the~monitoring point $m(x,y,z) = m(0,0,z_\mathrm{interface})$ in the~period of $t = \SI{0}{\second}$ to $\SI{10}{\second}$ were analysed utilising the~continuous wavelet transform. It~should be noted that no triggering action (such as welding current pulsation) was taken to excite melt-pool surface oscillations in the~present numerical simulations because even small surface fluctuations are detectable using the~proposed computational approach~\cite{Ebrahimi_2021}.

\begin{figure}[H] 
	\centering
	\includegraphics[width=0.9\linewidth]{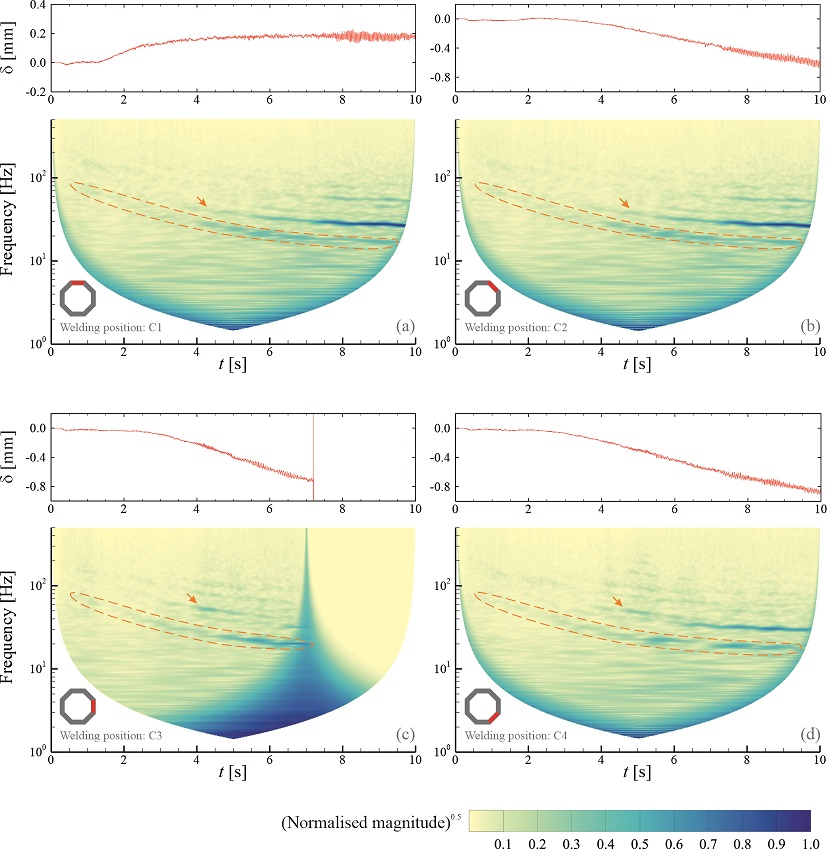}
	\caption{The~displacement signals recorded from the~monitoring point $m(x,y,z) = m(0,0,z_\mathrm{interface})$ and the~corresponding time-frequency spectra for different welding positions with downward welding direction. (a)~welding~positions~C1, (b)~welding~positions~C2, (c)~welding~positions~C3 and (d)~welding~positions~C4. Magnitudes are normalised with respect to the~maximum magnitude in the~time-frequency spectrum. Positive values of $\delta$ indicate surface depression and its negative values indicate surface elevation. ($I = \SI{85}{\ampere}$, travel~speed:~$\SI{2.5}{\milli\meter\per\second}$ and sulphur~concentration:~$\SI{240}{ppm}$)}
	\label{fig:frequency_position_down}
\end{figure}

\begin{figure}[H] 
	\centering
	\includegraphics[width=0.9\linewidth]{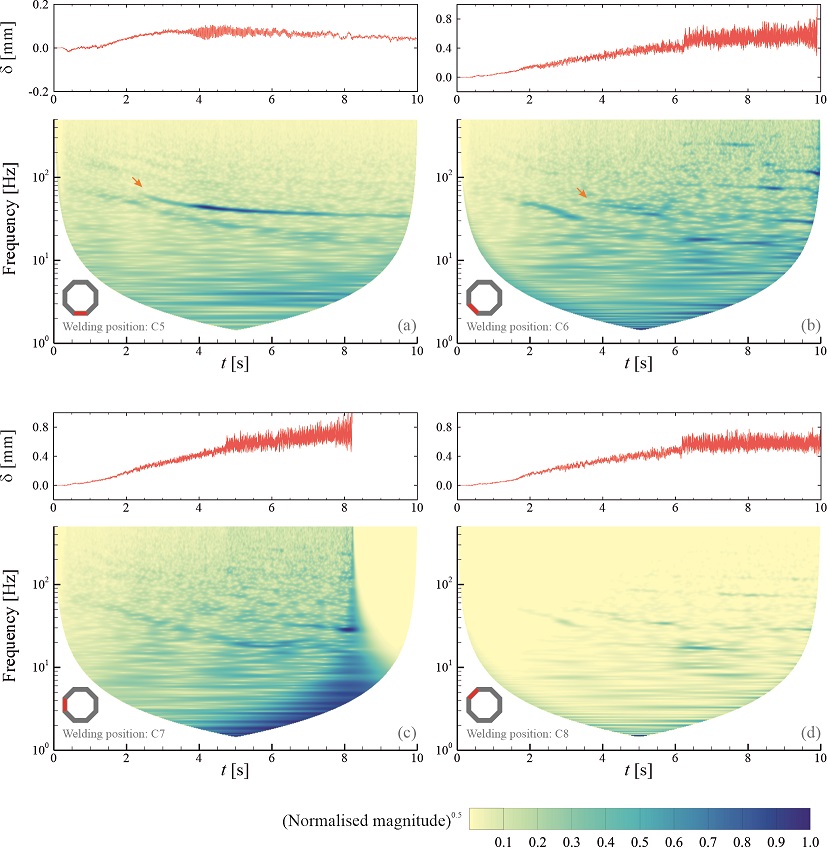}
	\caption{The displacement signals recorded from the monitoring point $m(x,y,z) = m(0,0,z_\mathrm{interface})$ and the~corresponding time-frequency spectrum for different welding positions with upward welding direction. (a)~welding~positions~C5, (b)~welding~positions~C6, (c)~welding~positions~C7 and (d)~welding~positions~C8. Magnitudes are normalised with respect to the~maximum magnitude in the time-frequency spectrum. Positive values of $\delta$ indicate surface depression and its negative values indicate surface elevation. ($I = \SI{85}{\ampere}$, travel~speed:~$\SI{2.5}{\milli\meter\per\second}$ and sulphur~concentration:~$\SI{240}{ppm}$)}
	\label{fig:frequency_position_up}
\end{figure}

The~oscillation signals of the~melt-pool surface and the~corresponding frequency spectra for different welding positions with $I = \SI{85}{\ampere}$ and sulphur concentration of $\SI{240}{ppm}$ are shown in \cref{fig:frequency_position_down} for downward welding direction (C1--C4) and in \cref{fig:frequency_position_up} for upward welding direction (C5--C8). For the~cases with a travel speed of $\SI{2.5}{\milli\meter\per\second}$, the~melt-pool depth increases over time and reaches the~plate thickness $H_\mathrm{m}$ in about $\SI{1.25}{\second}$. Then, the~melt-pool surface area on the~bottom surface increases and becomes almost the~same as that on the~top surface (\textit{i.e.} full penetration) at $t \approx \SI{4}{\second}$, after which the melt-pool grows to reach a~quasi-steady state. In the~case of GTA welding in position~C1 (\cref{fig:frequency_position_down}(a)), the~frequency of oscillations decreases from $f \approx \SI{73}{\hertz}$ at $t = \SI{1.25}{\second}$ to values of about $\SI{32}{\hertz}$ at $t = \SI{4}{\second}$, while the~melt-pool surface area on the~bottom surface is increasing to establish full penetration. After full penetration is established, the~frequency of the~most energetic event increases to $f \approx \SI{45}{\hertz}$, as indicated by arrow, and subsequently decreases gradually to values of about $\SI{32}{\hertz}$ as time passes and reaches $\SI{10}{\second}$. Both low and high frequencies remain in the~spectrum after full penetration. The~amplitudes of oscillations also augment as the~melt-pool size increases. The~frequency of oscillations obtained from the~present numerical simulations for welding position C1 agrees reasonably (within 10\% deviation bands) with the~experimental data reported by \mbox{Li~\textit{et al.}~\cite{Li_2018}}. The~range of oscillation frequencies predicted for different welding positions (C1--C8) seems to be the~same, however the~results suggest that the~welding position affects the~amplitudes of oscillations and the~evolution of melt-pool oscillatory behaviour. The~frequency of oscillations varies from $f \approx \SI{25}{\hertz}$ to values of about $\SI{37}{\hertz}$ after full penetration for welding positions C2~and~C4. Such increase in the~frequency of oscillations after full penetration also occurs for the~welding position C3, however the~frequency decreases after about $\SI{1}{\second}$ and the~melt pool oscillates at low frequencies up to $t \approx \SI{7}{\second}$, after which the~melt~pool collapses (\textit{i.e.} burns-through).

The~amplitude of oscillations predicted for cases C5--C8 are generally larger than those predicted for cases C1--C4, as shown in \cref{fig:frequency_position_up}. For the~welding position C5, the~frequency of oscillations decreases from $\SI{66}{\hertz}$ to $\SI{52}{\hertz}$ within $\SI{1}{\second}$ (from $t = \SI{1.25}{\second}$ to $t = \SI{2.25}{\second}$) and then the~frequency of oscillations increases to $\SI{64}{\hertz}$, as indicated by arrow. For~the~cases that welding direction is upward~(C6--C8), multiple changes occur in the~frequency of oscillations. The~abrupt changes observed in the~frequency domain indicate the~importance of utilising the~wavelet transform instead of the~Fourier transform for analysing the~behaviour of oscillating melt-pools. Changes in the~frequency of oscillations relate to changes in flow pattern in the~melt~pool, the~shape and size of the~melt-pool, and surface tension of the~molten metal \cite{Xiao_1993,Ebrahimi_2021}, which is discussed in \cref{sec:discussion}.

\begin{figure}[H] 
	\centering
	\includegraphics[width=0.9\linewidth]{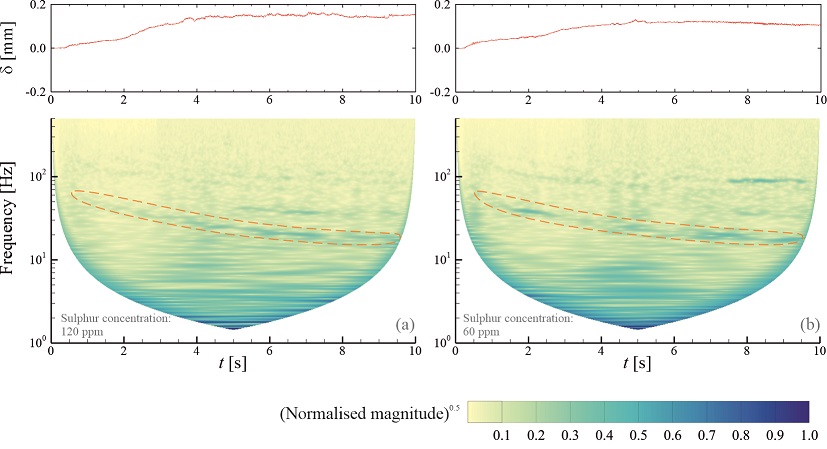}
	\caption{The~influence of sulphur concentration on the~time-frequency spectra of melt-pool oscillations during GTA~welding. (a)~sulphur concentration: $\SI{120}{ppm}$, and (b)~sulphur concentration: $\SI{60}{ppm}$ Magnitudes are normalised with respect to the~maximum magnitude in the~time-frequency spectrum. Positive values of $\delta$ indicate surface depression and its negative values indicate surface elevation. ($I = \SI{85}{\ampere}$, welding~position:~C1 and travel~speed:~$\SI{2.5}{\milli\meter\per\second}$)}
	\label{fig:frequency_sulphur}
\end{figure}

Changes in the~sulphur concentration of the~material, as a~surface-active element, can result in notable changes in surface tension of the~molten material and its variation with temperature ($\partial\gamma / \partial T$), as shown in \cref{fig:surface_tension}. Reducing the~amount of sulphur in the~material results in intensifying the~outward fluid flow over the~melt-pool surface (as shown in \cref{fig:meltpool_evolution_sc}), forming a~wide melt pool. The~effect of sulphur concentration of the~material on the~evolution of the~frequency of oscillations is shown in \cref{fig:frequency_sulphur} for welds in position~C1 and travel speed of $\SI{2.5}{\milli\meter\per\second}$. The~evolution of oscillation frequency is almost the~same for the~cases with sulphur concentrations of $\SI{120}{ppm}$ and $\SI{60}{ppm}$, and differs both qualitatively and quantitatively from that of the~case containing $\SI{240}{ppm}$ sulphur. The~melt pool oscillates at higher frequencies with larger amplitudes in the~case with sulphur concentration of $\SI{240}{ppm}$ compared to the~cases with sulphur concentrations of $\SI{120}{ppm}$ and $\SI{60}{ppm}$. Moreover, the~fundamental frequency of oscillations does not increase markedly after full penetration in the~cases with sulphur concentrations of $\SI{120}{ppm}$ and $\SI{60}{ppm}$, as it does in the~case with sulphur concentration of $\SI{240}{ppm}$. This difference in the~evolution of oscillations frequency relates to changes in the~structure of the~molten metal flow in the~melt pool and thus melt-pool shape evolution.

\begin{figure}[H] 
	\centering
	\includegraphics[width=0.9\linewidth]{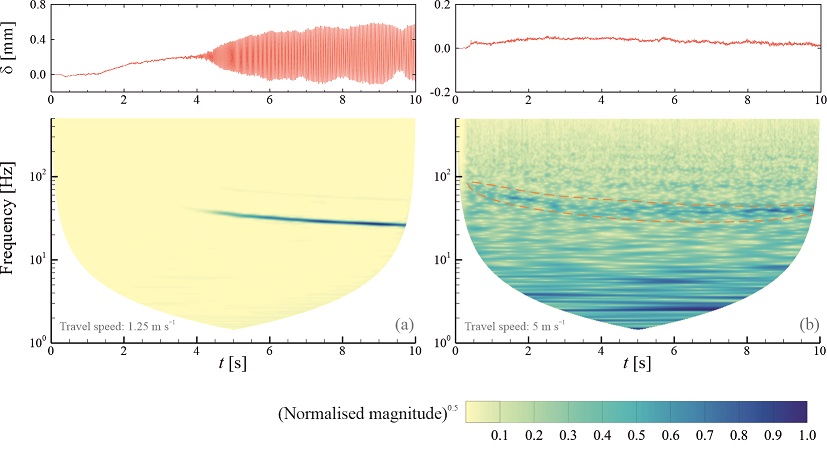}
	\caption{The~influence of welding travel speed on the~time-frequency spectra of melt-pool oscillations during GTA~welding. (a)~travel speed: $\SI{1.25}{\milli\meter\per\second}$, and (b)~travel speed: $\SI{5}{\milli\meter\per\second}$. Magnitudes are normalised with respect to the~maximum magnitude in the~time-frequency spectrum. Positive values of $\delta$ indicate surface depression and its negative values indicate surface elevation. ($I = \SI{85}{\ampere}$, welding~position:~C1 and sulphur~concentration:~$\SI{240}{ppm}$)}
	\label{fig:frequency_speed}
\end{figure}

The~influence of travel speed on the~frequency of melt-pool surface oscillations is shown in \cref{fig:frequency_speed} for the~welds in position~C1 and sulphur concentration of $\SI{240}{ppm}$. When the~travel speed was set to $\SI{1.25}{\milli\meter\per\second}$, the~melt pool reaches full penetration at $t \approx \SI{3}{\second}$ and the~amplitude of surface oscillations start to increase, as shown in \cref{fig:frequency_speed}(a). For the~case with a~travel speed of $\SI{1.25}{\milli\meter\per\second}$, the~frequency of oscillations gradually decreases from $\SI{69}{\hertz}$ at $t = \SI{1.25}{\second}$ to $\SI{29}{\hertz}$ at $t = \SI{10}{\second}$ while the~melt-pool is growing over time. Increasing the~travel speed from $\SI{1.25}{\milli\meter\per\second}$ to $\SI{5}{\milli\meter\per\second}$, the~amplitude of oscillations decreases significantly. Moreover, the~frequency of oscillations decreases up to $t = \SI{4}{\second}$ while the~melt-pool size is increasing. Afterwards, the~melt-pool reaches a~quasi-steady-state condition and the~variation of the~melt-pool size over time becomes insignificant, resulting surface oscillations at an almost constant frequency of about~$\SI{41}{\hertz}$. It should be noted that when the~pool size has reached steady state, the~surface area of the~melt pool on the~bottom surface is consistently smaller than that on the~top-surface in the~case with a~travel speed of $\SI{5}{\milli\meter\per\second}$.

\section{Discussion}
\label{sec:discussion}

\begin{figure}[H] 
	\centering
	\includegraphics[width=0.6\linewidth]{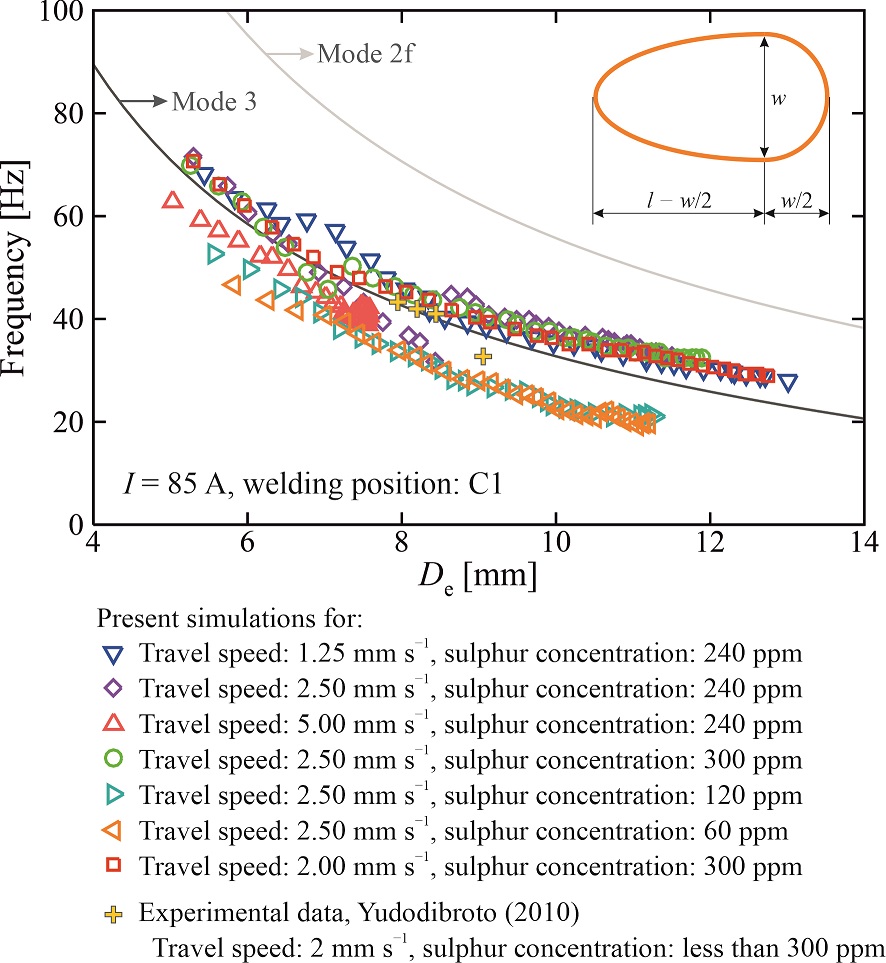}
	\caption{Variation of the~frequency of oscillations for fully-penetrated melt-pools during GTA welding as a~function of equivalent melt-pool~diameter. The~frequency of oscillations obtained form the~present numerical simulations (unfilled symbols), experimental data reported by Yudodibroto~\cite{Yudodibroto_2010_thesis} (squares), and analytical approximations using the~model proposed by Maruo and Hirata~\cite{Maruo_1993} for oscillation frequencies in Mode 3 (up~and~down~bulk~motion, dark~grey~line) and Mode 2f (sloshy~oscillation, light~grey~line). ($I = \SI{85}{\ampere}$ and welding~position:~C1)}
	\label{fig:frequency_analytical}
\end{figure}

The~frequency of oscillations predicted using the~present numerical simulations are compared to the~analytical approximations calculated from the~model developed by Maruo and Hirata~\cite{Maruo_1993} for fully-penetrated melt pools in welding position C1, and the~results are presented in \cref{fig:frequency_analytical}. This analytical model is expressed mathematically as follows~\cite{Maruo_1993}:

\begin{equation}
	f = \frac{1}{2\pi}\sqrt{\frac{2 \, \bar{\gamma} \, \mathscr{k}^2}{\rho \, H_\mathrm{m}} - \frac{2 \lVert\mathbf{g}\rVert}{H_\mathrm{m}}},
	\label{eq:analytical_frequency}
\end{equation}

\noindent
where, $H_\mathrm{m}$ is the~plate thickness, $\rho$ the~density of the~molten metal and $\bar{\gamma}$ the~average surface tension of the~molten material. The~mean value of surface tension for the~alloy considered in the~present study is approximately $\SI{1.6}{\newton\per\meter}$ in the~temperature range of $1723\text{--}\SI{2500}{\kelvin}$~\cite{Sahoo_1988}. The~value of $\mathscr{k}$ in \cref{eq:analytical_frequency} depends on both melt-pool size and oscillation mode and is obtained from the~following equations \cite{Yudodibroto_2010_thesis}: 

\begin{equation}
	\text{Mode 3: } \mathscr{k} = 2.405 {\left(\frac{D_\mathrm{e}}{2}\right)}^{-1},
	\label{eq:k_mode3}
\end{equation} 

\begin{equation}
	\text{Mode 2f: } \mathscr{k} = 3.832 {\left(\frac{D_\mathrm{e}}{2}\right)}^{-1},
	\label{eq:k_mode2f}
\end{equation} 

\noindent
where, $D_\mathrm{e}$ is the~equivalent diameter of the melt-pool under full penetration condition defined as follows:

\begin{equation}
	D_\mathrm{e} = \sqrt{l \, w},
	\label{eq:equivalent_diameter}
\end{equation} 

\noindent
where, $l$ and $w$ are the~melt-pool length and width on the~top surface respectively, as shown in \cref{fig:frequency_analytical}. The~analytical model of Maruo and Hirata~\cite{Maruo_1993} is developed for fully-penetrated melt pools, assuming that the~melt-pool shape and size on the~top and bottom surfaces of the~workpiece are the~same with no offset. It~appears that for the~case with a~travel speed of $\SI{2.5}{\milli\meter\per\second}$ and sulphur concentration of $\SI{240}{ppm}$ the~melt-pool surface oscillations follow the~analytically predicted frequencies in mode 3 for $t > \SI{4}{\second}$ when the~surface area of the~melt-pool on the~bottom surface approaches the~surface area on the~top surface. Reducing travel speed to $\SI{1.25}{\milli\meter\per\second}$ while keeping the~sulphur concentration unchanged~($\SI{240}{ppm}$), the~melt-pool surface area on the~bottom surface becomes almost the~same as on the~top surface and the~predicted frequencies follow the~analytical predictions in mode 3. For the~case with a~travel speed of $\SI{5}{\milli\meter\per\second}$ and sulphur concentration of $\SI{240}{ppm}$, the~melt pool shape and size reach a~quasi-steady-state with an equivalent diameter $D_\mathrm{e} \approx \SI{7.5}{\milli\meter}$ that does not change notably over time. In many practical applications, the~melt-pool surface area is not necessarily the~same on the~top and bottom surfaces of the~workpiece, particularly when the~melt pool is growing before reaching a~quasi-steady-state. Moreover, there is often an offset between the~positions of the~melt-pool surfaces on the~top and bottom of the~workpiece, which increases with increasing the~travel speed. These differences in the~melt-pool size and shape, as well as the~offset between the~melt-pool surfaces on the~top and bottom surfaces of the~workpiece, lead to deviation of the~frequencies approximated using the~analytical models from those predicted from numerical simulations and measured experimentally. The~frequencies predicted using the~present numerical simulations are in reasonably good agreement (within 10\% deviation bands) with the~experimental measurements of fully penetrated pools reported by~\mbox{Yudodibroto~\cite{Yudodibroto_2010_thesis}}.

Despite the~suitability of analytical models for predicting the~frequency of oscillations under a~full penetration condition, they fail to predict changes in oscillation mode during welding processes, particularly when the~melt pool is evolving over time. Moreover, variation in the~value of surface tension of the~molten material with temperature is ignored in the~analytical models, which limits their accuracy in predicting the~frequency of oscillations during GTA welding. The~results presented in \cref{fig:frequency_analytical} also demonstrate that changes in the~sulphur~concentration can affect the~frequency of oscillations and their evolution due to variations in the~internal flow pattern and thus the~melt-pool shape, which results from changes in the~Marangoni stresses acting on the~melt-pool surfaces.

\begin{figure}[H] 
	\centering
	\includegraphics[width=0.8\linewidth]{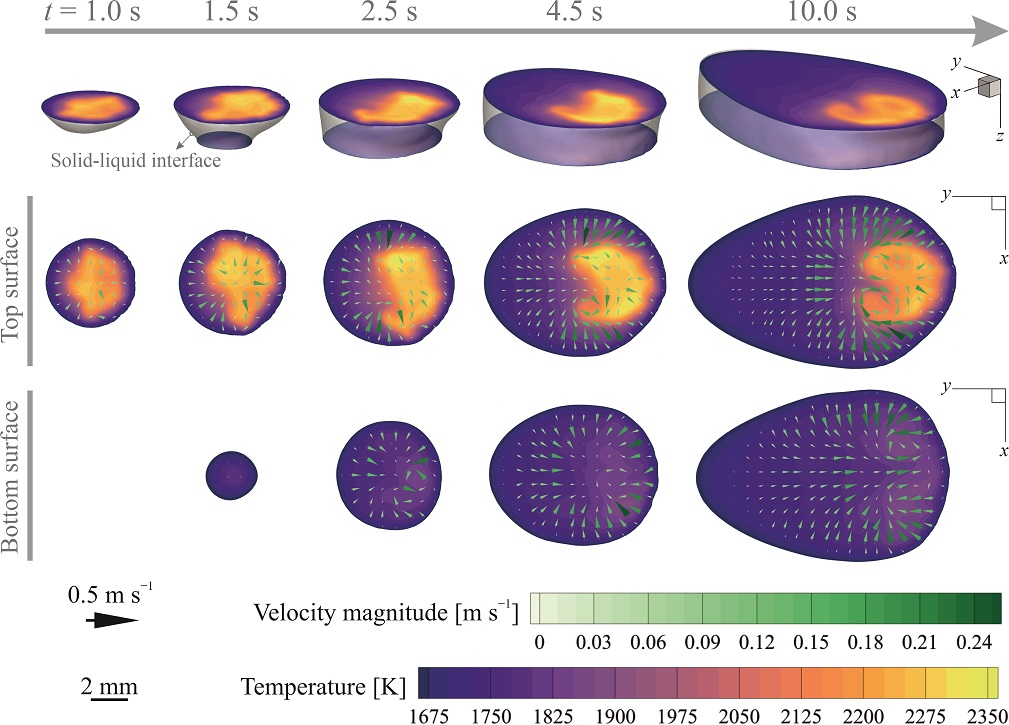}
	\caption{Evolution of the~melt-pool shape during gas tungsten arc welding. Contours show the~temperature distribution over the~melt-pool surface and are overlaid with velocity vectors. The~$\SI{0.5}{\meter\per\second}$ reference vector is provided for scaling the~velocity field. ($I = \SI{85}{\ampere}$, welding~position:~C1, travel speed: $\SI{2.5}{\milli\meter\per\second}$ and sulphur~concentration:~$\SI{240}{ppm}$)}
	\label{fig:meltpool_evolution_c1}
\end{figure}

\Cref{fig:meltpool_evolution_c1} shows the~evolution of the~thermal and flow fields over the~melt-pool surfaces as well as the~pool shape during GTA welding in position C1 with a~travel speed of $\SI{2.5}{\milli\meter\per\second}$ and sulphur concentration~of~$\SI{240}{ppm}$. A~melt pool forms soon after the~arc ignition, grows and its depth reaches the~plate thickness after about $\SI{1.25}{\second}$, forming a~fully-penetrated melt pool. Fluid flow in the~melt pool is driven by various time-variant forces acting on the~molten material such as Marangoni, Lorentz, arc plasma shear and pressure and buoyancy forces, resulting in a~complex flow pattern that is inherently three-dimensional. This fluid motion transfers the~heat absorbed by the~material and affects the~melt-pool shape and its evolution over time. The~relative contribution of advective to diffusive energy transfer can be evaluated using the~P{\'e}clet number ($\mathrm{Pe} = \rho c_\mathrm{p} \mathscr{D} \lVert \mathbf{u} \rVert / k$), which is larger than unity ($\mathcal{O}(100)$) for the~cases studied in the~present work, signifying the~notable influence of advection on the~melt-pool shape.

The~results shown in \cref{fig:meltpool_evolution_c1} indicate that the~maximum temperature of the~melt pool after reaching a~quasi-steady-state condition varies between $\SI{2260}{\kelvin}$ and $\SI{2340}{\kelvin}$ and the~maximum local fluid velocity varies between $\SI{0.21}{\meter\per\second}$ and $\SI{0.34}{\meter\per\second}$. Inward fluid flow from the~boundary to the~centre of the~pool is observed over the~top surface that meets an outward flow in the~central region. This change in the~flow direction is due to the~sign change of the~temperature gradient of surface tension ($\partial\gamma / \partial T$) at a~specific temperature, as shown in \cref{fig:surface_tension}. Interactions between these two streams disturb the~thermal field over the~pool surface and generate an unsteady complex flow pattern inside the~pool, affecting the~energy transport in the~melt pool and thus the~evolution of the~melt-pool shape. Temperatures are below the~critical temperature over the~bottom surface of the~pool and the~temperature gradient of the~surface tension is positive all over the~surface, resulting in inward fluid flow from the~melt-pool boundaries. Moreover, two vortices form over the~top surface as time passes that generate a~periodic asymmetry in the~flow field, leading to flow oscillations around the~melt-pool centreline. The~formation of such vortices occurs because of the~fluid motion from the~front part of the~pool toward the~rear and collision with the~inward flow in the~rear part of the~pool. A~similar flow pattern was observed during GTA welding by~\mbox{Zhao~\textit{et al.}~\cite{Zhao_2009}} using particle image velocimetry~(PIV).

The~outward fluid flow in the~central region of the~melt pool coupled with the~arc pressure applied to the~molten material leads to melt-pool surface depressions in the~front and central region of the~melt pool. Variation in the~flow pattern over time, as well as changes in the~melt-pool shape result in changes in oscillatory behaviour. For the~case shown in \cref{fig:meltpool_evolution_c1}, the~melt-pool surface area on the~bottom surface of the~workpiece increases over time and becomes almost the~same as that on the~top surface after $\SI{4}{\second}$. This change in pool shape coupled with the~fluid flow that evolves over the~bottom surface, causes a~change in the~melt-pool oscillatory behaviour, as reflected in \cref{fig:frequency_position_down}. Variations in the~pool surface morphology also result in variation of the~power-density and arc force distribution over the~surface and the~total power input from the~electric arc, enhancing flow disturbances. 

\begin{figure}[H] 
	\centering
	\includegraphics[width=0.9\linewidth]{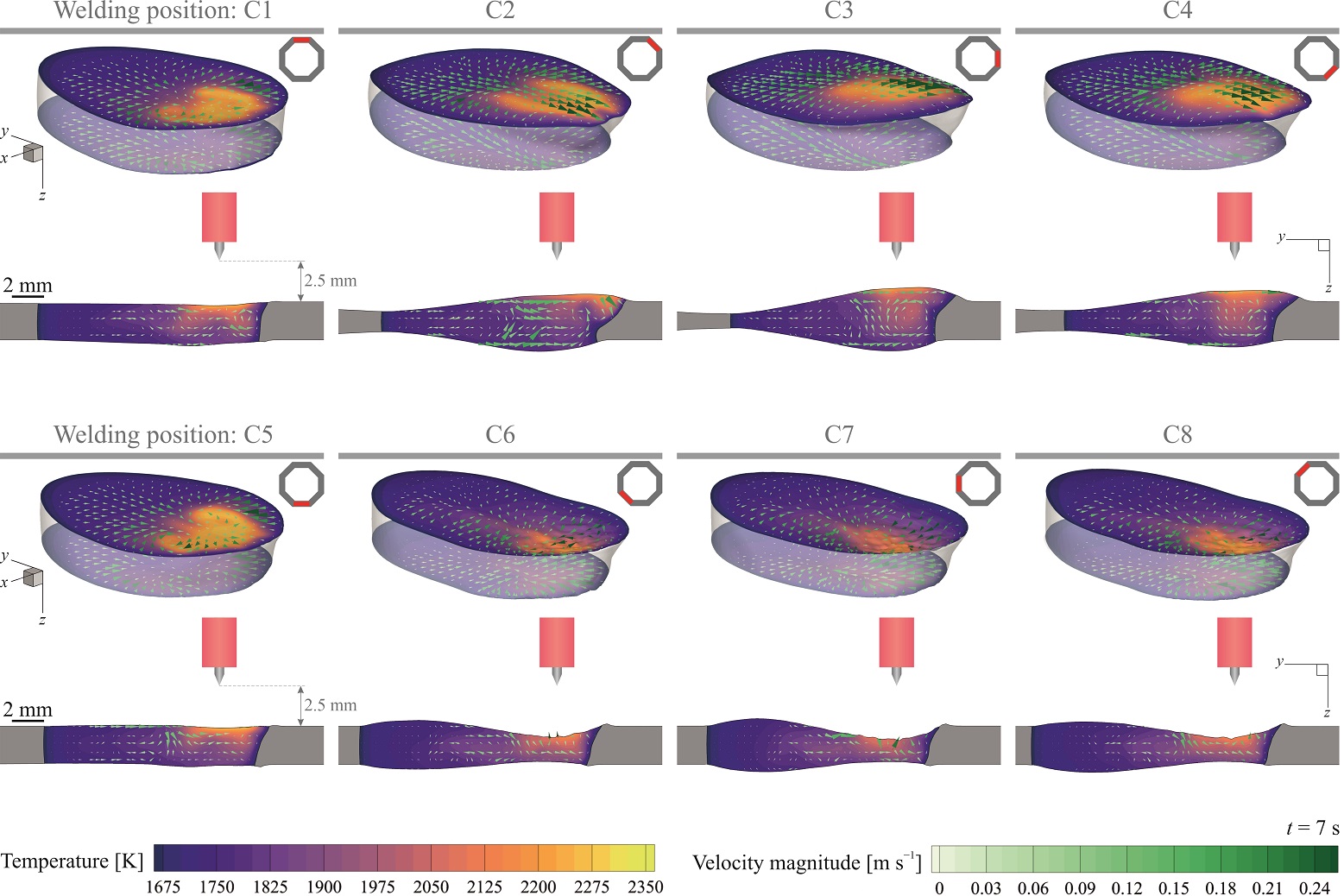}
	\caption{The~influence of the~welding position on evolution of the~melt-pool shape during gas tungsten arc welding. Contours show the~temperature profile at $t = \SI{7}{\second}$ and are overlaid with velocity vectors. Cross sections on the~$y$-$z$ plane are located at $x = 0$. ($I = \SI{85}{\ampere}$, travel speed: $\SI{2.5}{\milli\meter\per\second}$ and sulphur~concentration:~$\SI{240}{ppm}$)}
	\label{fig:meltpool_evolution_wp}
\end{figure}

Variation in the~welding position can affect melt-pool surface deformations, resulting in changes in the~power-density and arc force distribution over the~melt-pool surface and thus the~evolution of the~melt-pool shape and its oscillatory behaviour. The~influence of welding position on melt-pool shape is shown in \cref{fig:meltpool_evolution_wp} for cases with a~travel~speed~of~$\SI{2.5}{\milli\meter\per\second}$ and sulphur~concentration of~$\SI{240}{ppm}$. When welding downward~(C2--C4), the~molten material is pulled by the~gravitational force towards the~front part of the~melt pool and forms a~bulge beneath the~welding torch. This change in the~melt-pool surface morphology decreases the~average arc length and alters the~power-density distribution, through changing the~distribution parameter $\sigma_\mathrm{q}$ (\cref{eq:dist_heat}), and thus temperature profile over the~surface. The~change in the~power-density distribution in turn increases the~temperature gradients over the~melt-pool surface and thus the~magnitude of Marangoni forces. The~results reveal that the~molten material moves from the~rear of the~melt pool towards the~front around the~centreline, resulting in further reduction of the~melt-pool thickness in the~rear part of the~melt pool, which can eventually lead to the~rupture of liquid layer and melt-pool collapse if the~dynamic force balance cannot be maintained. In contrast, when welding upward (C6--C8), the~molten material moves towards the~rear part of the~melt pool because of the~gravitational force, forming a~concavity in the~front part of the~pool. The~formation of this concavity increases the~average arc length beneath the~welding torch and the~distribution parameter $\sigma_\mathrm{q}$ (\cref{eq:dist_heat}). This increase in the~distribution parameter reduces the~temperature gradients over the~melt pool surface and thus the~magnitude of Marangoni forces. These results suggest that although the~melt-pool shape and its surface morphology are influenced by the~welding position, the~overall flow structure in the~melt pool, which is dominated by Marangoni and electromagnetic forces, is not affected significantly by the~gravitational force~\cite{Nguyen_2017}. The~results in \cref{fig:meltpool_evolution_wp} also show that the~material thickness~$H$ reduces locally beneath the~welding torch with the~increase in the~average arc length in the~cases that welding direction is upward (C6--C8). This reduction in the~material thickness results in the~establishment of full penetration somewhat earlier at $t \approx \SI{3}{\second}$ compared to that of the~case C1 and C5, changing the~oscillation frequency as reflected in \cref{fig:frequency_position_up}. The~results suggest that when the~relative material thickness ($H / H_\mathrm{m}$) beneath the~welding torch reduces to values less than about 0.65, an~unsteady multicellular flow pattern evolves in the~pool~\cite{Ebrahimi_2021,Schatz_2001}, leading to irregular surface deformations that are reflected in the~time-frequency spectra shown in \cref{fig:frequency_position_up}(b and d) for $t > \SI{6}{\second}$.

\begin{figure}[H] 
	\centering
	\includegraphics[width=0.8\linewidth]{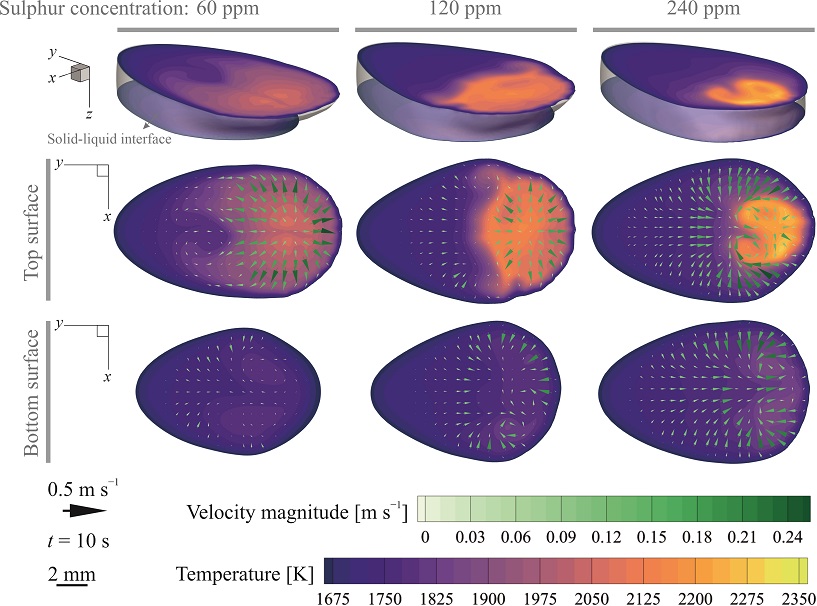}
	\caption{The~influence of sulphur concentration in the~material on the~melt-pool shape during gas tungsten arc welding. Contours show the~temperature profile at $t = \SI{10}{\second}$ and are overlaid with velocity vectors. The~$\SI{0.5}{\meter\per\second}$ reference vector is provided for scaling the~velocity field. ($I = \SI{85}{\ampere}$, travel speed: $\SI{2.5}{\milli\meter\per\second}$ and welding position: C1)}
	\label{fig:meltpool_evolution_sc}
\end{figure}

\Cref{fig:meltpool_evolution_sc} shows the~melt-pool shape and thermal and flow fields over the~melt-pool surfaces at $t = \SI{10}{\second}$ for GTA welding in welding position C1 with a~travel speed of $\SI{2.5}{\milli\meter\per\second}$ and different sulphur concentrations in the~material. The~results of the~present numerical simulations suggest that both the~amplitude and frequency of oscillations decrease with reducing the~sulphur concentration in the~material. Reducing sulphur concentration in the~material results in an increase in the~average surface tension of the~molten material, affects the~variation of surface tension with temperature ($\partial\gamma / \partial T$) and reduces the~critical temperature at which the~sign of the~temperature gradient of surface tension ($\partial\gamma / \partial T$) changes, according to \cref{eq:surface_tension} (see \cref{fig:surface_tension}). The~flow pattern over the~melt-pool top surface becomes mostly outward when the~sulphur concentration in the~material is reduced. The~increasingly outward fluid flow weakens disturbances caused by the~interaction of inward and outward fluid flows over the~melt-pool surface. Additionally, the~outward fluid flow transfers the~heat absorbed by the~material towards the~melt-pool boundary, reducing the~maximum temperature and the~magnitude of temperature gradients over the~melt-pool surface. Moreover, the~outward flow results in a~relatively larger melt-pool surface area on the~top surface compared to that on the~bottom surface and fluid velocities decrease on the~bottom surface. These changes in the~flow field alter the~melt-pool shape, as shown in \cref{fig:meltpool_evolution_sc}, and in turn affects the~melt-pool oscillatory behaviour. A notable change in the~time-frequency spectra of the~cases with sulphur concentrations of $\SI{120}{ppm}$ and $\SI{60}{ppm}$ (\cref{fig:frequency_sulphur}) compared to that with a~sulphur concentration of $\SI{240}{ppm}$ (\cref{fig:frequency_position_down}(a)) is that the~frequency of oscillations decreases gradually over time and does not change suddenly as observed in \cref{fig:frequency_position_down}(a) at $t \approx \SI{4}{\second}$.

\begin{figure}[H] 
	\centering
	\includegraphics[width=0.8\linewidth]{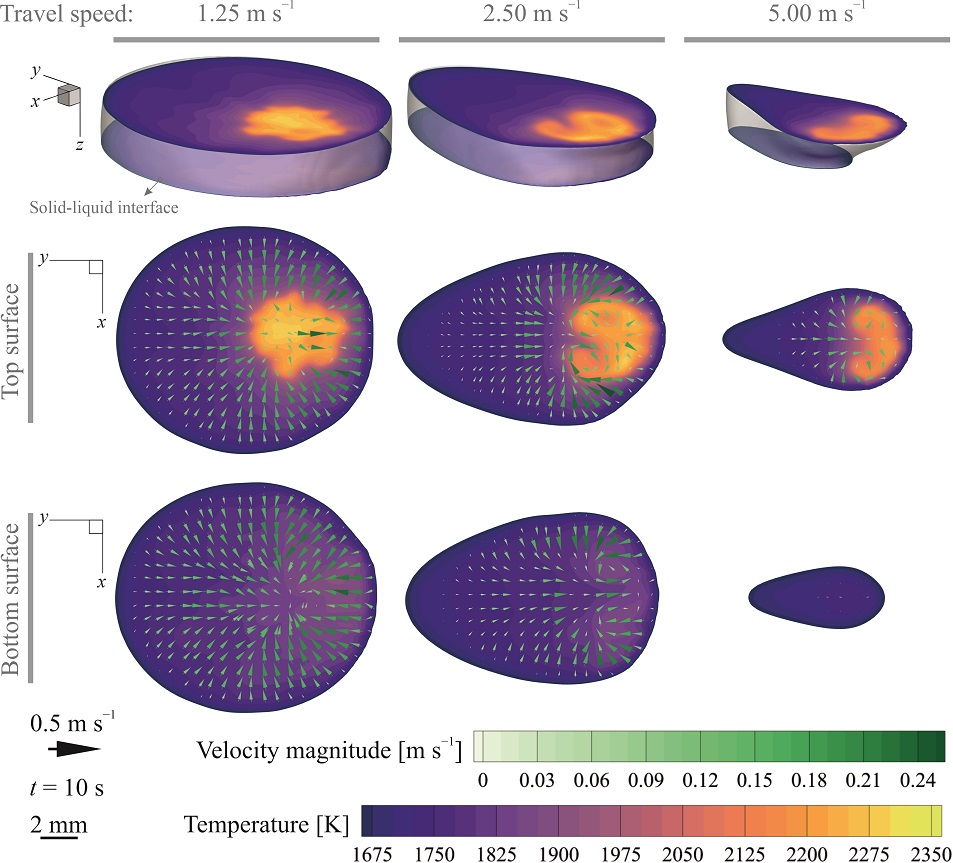}
	\caption{The~influence of travel speed on the~melt-pool shape during gas tungsten arc welding. Contours show the~temperature profile at $t = \SI{10}{\second}$ and are overlaid with velocity vectors. The~$\SI{0.5}{\meter\per\second}$ reference vector is provided for scaling the~velocity field. ($I = \SI{85}{\ampere}$, sulphur concentration: $\SI{240}{ppm}$ and welding position: C1)}
	\label{fig:meltpool_evolution_ts}
\end{figure}

The~influence of travel speed on melt-pool shape and heat and fluid flow in the~melt pool during GTA welding in position C1 is shown in \cref{fig:meltpool_evolution_ts} for the~cases with a~sulphur~concentration of~$\SI{240}{ppm}$. Increasing the~travel speed, while keeping other process parameters the~same, results in a~decrease in the~melt-pool size, which is due to the~reduction of nominal heat input to the~material. Additionally, the~melt-pool shape changes from a~virtually circular shape to a~teardrop shape with increasing travel speed. The~flow pattern obtained from the~numerical simulations also reveals that vortex structures do not form over the~melt-pool surface for the~case with a~travel speed of $\SI{1.25}{\milli\meter\per\second}$, in contrast to other cases with higher travel speeds. Moreover, the~offset between the~top and bottom melt-pool surfaces increases with increasing the~travel speed. The~offset measured at the~leading edge of the~pool with respect to the~$z$-axis increases from about $\ang{5}$ at a~travel speed of $\SI{1.25}{\milli\meter\per\second}$ to values of about $\ang{50}$ at a~travel speed of $\SI{5}{\milli\meter\per\second}$.

\section{Conclusions}
\label{sec:conclusions}

High-fidelity three-dimensional numerical simulations were performed to study the~oscillatory behaviour of fully-penetrated melt pools during positional gas tungsten arc welding. The~influence of welding positions, the~sulphur concentration in the~material and travel speed on complex unsteady convection in the~melt pool and oscillations of the~melt-pool surface were investigated. The~frequencies predicted using the~present computational model are compared with analytical and experimental data, and reasonably good agreement (within 5\% deviation bands) is achieved. Using the~present numerical approach, evolutions of the~melt-pool surface oscillations during GTA welding are described by revealing the~unsteady complex flow pattern in the~melt pools and subsequent changes in the~melt-pool shape, which are generally difficult to visualise experimentally. Moreover, evolution of the~frequency of melt-pool oscillations during arc welding are not predictable using the~analytical models that are available in the~open literature.

Melt-pool oscillatory behaviour depends on surface tension of the~molten material and shape and size of the~melt-pool. Changes in material properties and welding process parameters affect convection in the~melt pool, hydrodynamic instabilities that arise and resultant variations in the~melt-pool shape. Depending on the~processing condition, these instabilities can also grow in time, affecting melt-pool stability and may even lead to melt-pool collapse and process failure. Welding position affects the~melt-pool surface morphology, altering the~spatial distribution of arc forces and power-density applied to the~molten material and thus changes flow pattern in the~melt-pool. The~change in the~flow pattern affects the~evolution of the~melt-pool shape and its oscillatory behaviour. The~frequency of oscillations seems to vary within the~same range ($\SI{22}{\hertz} < f < \SI{73}{\hertz}$) for different welding positions studied in the~present work, however the~evolution of oscillation frequencies, which depends on the~melt-pool shape, is affected by welding position. Under similar welding conditions, sulphur concentration in the~material significantly affects the~thermal and flow fields in the~melt pool and consequently the~shape of the~pool, changing the~oscillatory behaviour of the~melt-pool. Increasing the~travel speed decreases the~melt-pool size, increases the~offset between top and bottom melt-pool surfaces and also affects the~flow structures (vortex formation) on the~melt-pool surface. These observations offer an insight into the~complex melt-pool oscillatory behaviour during positional gas tungsten arc welding and suggest that the~processing window for advanced fusion-based manufacturing processes can be determined by utilising numerical simulations that can potentially reduce the~costs associated with process development and optimisation.

\section*{Acknowledgement}
\label{sec:acknowledgement}

This research was carried out under project number F31.7.13504 in the~framework of the~Partnership Program of the~Materials innovation institute M2i (www.m2i.nl) and the~Foundation for Fundamental Research on Matter (FOM) (www.fom.nl), which is part of the~Netherlands Organisation for Scientific Research (www.nwo.nl). The~authors would like to thank the~industrial partner in this project “Allseas Engineering B.V.” for the~financial support.

\section*{Author Contributions}
\label{sec:author_contributions}

Conceptualisation, A.E., C.R.K. and I.M.R.; methodology, A.E.; software, A.E.; validation, A.E.; formal analysis, A.E.; investigation, A.E.; resources, A.E., C.R.K, M.J.M.H., and~I.M.R.; data curation, A.E.; writing---original draft preparation, A.E.; writing---review and editing, A.E., C.R.K., M.J.M.H., and~I.M.R; visualisation, A.E.; supervision, C.R.K. and I.M.R.; project administration, A.E. and I.M.R.; and funding acquisition, I.M.R.

\section*{Conflict of interest}
\label{sec:conflict_of_interest}

The authors declare no conflict of interest.

\section*{Data availability}
\label{sec:data_availability}

The~raw/processed data required to reproduce these findings cannot be shared at this time due to their large size, but representative samples of the~research data are presented in the~paper. Other datasets generated during this study are available from the~corresponding author on reasonable request.

\small{
	\bibliographystyle{elsarticle-num}
	\bibliography{ref}
}

\end{document}